\documentclass[conference]{IEEEtran}
\IEEEoverridecommandlockouts
\usepackage{cite}
\usepackage{amsmath,amssymb,amsfonts}
\usepackage{algorithmic}
\usepackage{graphicx}
\usepackage{textcomp}
\usepackage{xcolor}
\usepackage{verbatim}
\def\BibTeX{{\rm B\kern-.05em{\sc i\kern-.025em b}\kern-.08em
    T\kern-.1667em\lower.7ex\hbox{E}\kern-.125emX}}
\begin{document}

\title{Learning to Unfold Fractional Programming for Multi-Cell MU-MIMO Beamforming \\with Graph Neural Networks\\
}

\author{\IEEEauthorblockN{Zihan Jiao, Xinping Yi, Shi Jin}
\IEEEauthorblockA{School of Information Science and Engineering, Southeast University, Nanjing 210096, China\\
\{230248863, xyi, jinshi\}@seu.edu.cn
}
}

\maketitle

\begin{abstract}
In the multi-cell multiuser multi-input multi-output (MU-MIMO) systems, fractional programming (FP) has demonstrated considerable effectiveness in optimizing beamforming vectors, yet it suffers from high computational complexity. Recent improvements demonstrate reduced complexity by avoiding large-dimension matrix inversions (i.e., FastFP) and faster convergence by learning to unfold the FastFP algorithm (i.e., DeepFP).
However, the state-of-the-art DeepFP algorithm relies heavily on FastFP with
not-fast-enough convergence and suffers from limited generalization capability. To address these issues, this paper proposes to further accelerate DeepFP through learning to directly unfold the classical FP algorithm with a novel structure-aware graph neural network (GNN) model.
The proposed algorithm, named GNNFP, first 
reformulates the beamforming optimization within the FP alternating iteration framework into a standard single-variable quadratic form, which can be efficiently solved by the GNN model.
Numerical experiments show that the proposed algorithm achieves competitive performance while offering substantial advantages over other FP-based algorithms in terms of model size, computational time, and generalization capability.


\end{abstract}

\section{Introduction}
Multi-cell MU-MIMO has become a cornerstone technology for next-generation wireless networks, enabling significant improvements in spectral efficiency and network capacity through coordinated transmission and interference management \cite{MU-MIMO}. In such systems, beamforming plays a critical role in exploiting spatial degrees of freedom while mitigating inter-cell and inter-user interference. 
Naturally, optimizing beamforming design to maximize the weighted sum-rate (WSR) has become one of the most fundamental and critical problems in MU-MIMO systems. 
However, the coupled interference terms and power constraints introduce high non-convexity, making the problem extremely challenging to solve.

The general WSR problem is NP-hard. Although linear precoding methods, such as maximum ratio transmission (MRT) and zero-forcing (ZF), seem appealing from a practical point of view, there are unavoidable performance losses. Recently, the weighted minimum mean square error (WMMSE) algorithm has been widely applied, demonstrating excellent balance between performance and computational complexity. The WSR problem can be rewritten as a weighted MSE minimization problem and consequently solved by the block coordinate descent (BCD) method in an iterative manner. 

Most recently, with the rapid development of artificial intelligence technology, learning-based approaches have demonstrated increasingly substantial contributions to solving WSR problems. 
One major paradigm is the model-driven methods that adopt the deep unfolding technique, where neural networks are utilized to replace certain components of conventional optimization algorithms. Among them, \cite{IAIDNN} introduced  trainable matrices that approximate matrix inversion to unfold WMMSE, though limited to single-cell scenarios; \cite{GCN-WMMSE} proposed a GCN-WMMSE architecture that employs graph convolutional structures to replace the update steps for both the receiver weight matrix and beamforming vectors in the original WMMSE algorithm, which could significantly reduce the iterations; \cite{UWMMSE} also unfolded the receiver weight matrix update step of the original WMMSE algorithm, distinctively employing dual neural architectures (GNN and MLP) to leverage their respective strengths. Another thread of research promotes data-driven approaches, where, for example, \cite{ENGNN} proposed a novel edge-update mechanism followed by an edge-update empowered neural network architecture termed edge-node graph neural network (ENGNN), which collaboratively maps the problem parameters into the proposed GNN framework for training and inference; \cite{Early Access} reformulated the prime problem into a reduced-dimensional space that preserves the solution invariance property, subsequently employing a convolutional neural network (CNN) for solutions. 

Besides the aforementioned methodologies, fractional programming (FP)-based algorithms have progressively matured into a comprehensive and robust technical framework to tackle the WSR problem, exhibiting superior performance characteristics.
The classical FP algorithm \cite{FPPart1}, \cite{FPPart2}, which is mathematically equivalent to the well-known WMMSE algorithm \cite{WMMSE}, guarantees convergence to stationary points but suffers from high computational complexity due to bisection search for power constraint multipliers, where each evaluation requires matrix inversion.
To mitigate this bottleneck, FastFP \cite{FastFP} introduces a nonhomogeneous transform that eliminates matrix inversions, at the cost of slow convergence due to conservative stepsize selection based on the coupling matrix's spectral properties. 
More recently, DeepFP \cite{DeepFP} leverages deep unfolding \cite{Unfolding tech} to learn adaptive stepsizes, achieving faster convergence than FastFP. However, it employs layer-specific parameters and consequently exhibits limited generalization across different problem sizes and iteration numbers.

To overcome the slow convergence of FastFP and the limited generalization capability of DeepFP, this paper proposes GNNFP, a novel algorithm that directly unfolds the classical FP framework using structure-aware GNN. The main contributions and novelties are as follows: First, through a sequence of equivalent transformations, we reformulate the beamforming update subproblem within the classical FP framework into a standard single-variable quadratic form, making it amenable to graph-based learning. Second, inspired by the success of structure-aware GNN in solving combinatorial optimization problems \cite{PIGNN}, we design a GNN architecture that directly reflects the quadratic structure of the reformulated problem, with shared parameters across iterations that enable superior generalization capability across different iteration numbers and problem sizes, thereby fundamentally overcoming DeepFP's limitations. Third, our methodology provides a general learning-based framework for solving constrained multi-variable quadratic programming problems. Numerical results demonstrate that GNNFP achieves competitive performance while offering substantial advantages in convergence speed, computational efficiency, model compactness, and generalization capability compared to existing FP-based algorithms.

\section{Problem Formulation}
Consider an $L$-cell downlink multicell MU-MIMO network which has $Q$ users in each cell. Each BS and each UE are equipped with  $N_{t}$  and  $N_{r}$  antennas, respectively. In the additive white Gaussian noise (AWGN) channel with noise power $\sigma^2_{N}$, the received signal is given by
\begin{equation}\small
\begin{aligned}
\mathbf{y}_{\ell q}&=\underbrace{\mathbf{H}_{\ell q,\ell}\mathbf{v}_{\ell q}\mathrm{s}_{\ell q}}_{\text{desired signal}}+\underbrace{\sum_{j=1,j\neq q}^Q\mathbf{H}_{\ell q,\ell}\mathbf{v}_{\ell j}\mathrm{s}_{\ell j}}_{\text{intracell interference}}\\[-1pt]
&+\underbrace{\sum_{i=1,i\neq\ell}^L\sum_{j=1}^Q\mathbf{H}_{\ell q,i}\mathbf{v}_{ij}\mathrm{s}_{ij}}_{\text{intercell interference}}+\mathbf{n}_{\ell q},
\end{aligned}
\end{equation}
where $\mathbf{H}_{\ell q,i}\in\mathbb{C}^{N_{r}\times N_t}$ is the channel matrix from BS ${i}$ to user $(\ell,q)$, $\mathbf{v}_{\ell q}\in\mathbb{C}^{N_{t}}$ denotes the beamforming vector from BS ${\ell}$ to user $(\ell,q)$.
By treating interference as noise (TIN) at the users, the achievable data rate for user $(\ell,q)$ is given as 
\begin{equation}\mathrm{R}_{\ell q}=\log(1+\mathbf{v}_{\ell q}^H\mathbf{H}_{\ell q,\ell}^H\mathbf{C}_{\ell q}^{-1}\mathbf{H}_{\ell q,\ell}\mathbf{v}_{\ell q}),\end{equation}
where $\mathbf{C}_{\ell q}\in\mathbb{C}^{N_{t}\times N_t}$ represents the interference-plus-noise covariance matrix at user $(\ell,q)$ which could be expanded as
\begin{equation}\begin{aligned}\mathbf{C}_{\ell q}&=\sum_{j=1,j\neq q}^Q\mathbf{H}_{\ell q,\ell}\mathbf{v}_{\ell q}\mathbf{v}_{\ell q}^H\mathbf{H}_{\ell q,\ell}^H\\&+\sum_{i=1,i\neq\ell}^L\sum_{j=1}^Q\mathbf{H}_{\ell q,i}\mathbf{v}_{ij}\mathbf{v}_{ij}^H\mathbf{H}_{\ell q,i}^H+\sigma^2_{N}\mathbf{I}_{N_{r}}.\end{aligned}\end{equation}

We aim to optimize the set of beamforming vectors $\underline{\boldsymbol{v}} = \{\mathbf{v}_{\ell q}: \ell \in \{1,\ldots,L\}, q \in \{1,\ldots,Q\}\}$ under the transmit power constraint of each cell, so as to maximize the weighted sum-rate throughout the network, which could be mathematically formulated as
\begin{subequations}\label{origin_problem}
\begin{align}
\max_{\underline{\boldsymbol{v}}}\quad&\sum_{\ell=1}^L\sum_{q=1}^Q\omega_{\ell q}\mathrm{R}_{\ell q} \label{origin_problem_obj}\\
\mathrm{s.t.}\quad&\sum_{q=1}^Q\|\mathbf{v}_{\ell q}\|_2^2\leq {P}_\ell,\quad\ell=1,\ldots,L \label{origin_problem_const}
,\end{align}
\end{subequations}
where $\omega_{\ell q}$ represents the non-negative weight coefficient for user $({\ell, q})$, the constraint (\ref{origin_problem}b) defines that the total transmit power must comply with the power budget $P_{\ell}$ of the BS $\ell$.

\section{FP-Based Algorithms}
Addressing the problem (\ref{origin_problem}) formulated in Section II, FP-based algorithms have emerged as a well-established methodology with provable convergence and strong empirical performance. This section reviews three FP-based algorithms that represent a progressive evolution in addressing computational efficiency: the classical FP algorithm that establishes the theoretical foundation, the FastFP algorithm that eliminates large matrix inversions to reduce complexity, and the DeepFP algorithm that further leverages deep unfolding to accelerate convergence through learned parameters. 
\subsection{Classical FP Algorithm~\cite{FPPart1,FPPart2}}

To solve the constrained optimization problem (\ref{origin_problem}), the classical FP algorithm applies Lagrangian dual transform and quadratic transform successively to reformulate the objective into an unconstrained form $f_t(\underline{\boldsymbol{v}}, \underline{\boldsymbol{y}}, \underline{\boldsymbol{\gamma}})$ that is concave with respect to each variable block
\begin{equation}
\begin{aligned}
f_t(\underline{\boldsymbol{v}}, \underline{\boldsymbol{y}}, \underline{\boldsymbol{\gamma}}) 
&= \sum_{\ell=1}^{L} \sum_{q=1}^{Q} \Big[ 2\Re\{\omega_{\ell q}(1 + \gamma_{\ell q})\mathbf{v}_{\ell q}^\mathit{H} \mathbf{H}_{\ell q, \ell}^\mathit{H} \mathbf{y}_{\ell q}\} \\
&\quad - \mathbf{v}_{\ell q}^\mathit{H} \mathbf{D}_\ell \mathbf{v}_{\ell q} - \omega_{\ell q}(1 + \gamma_{\ell q})\sigma_{N}^2 \mathbf{y}_{\ell q}^\mathit{H} \mathbf{y}_{\ell q} \\
&\quad + \omega_{\ell q} \log(1 + \gamma_{\ell q}) - \omega_{\ell q}\gamma_{\ell q} \Big],
\end{aligned}
\end{equation}
where 
$$\mathbf{D}_\ell=\sum_{i=1}^L\sum_{j=1}^Q\omega_{ij}(1+\gamma_{ij})\mathbf{H}_{ij,\ell}^\mathit{H}\mathbf{y}_{ij}\mathbf{y}_{ij}^\mathit{H}\mathbf{H}_{ij,\ell}$$
with
$\underline{\boldsymbol{y}} = \{\mathbf{y}_{\ell q}\}$ and 
$\underline{\boldsymbol{\gamma}}=\{\boldsymbol{\gamma}_{\ell q}\}$ 
for $\ell \in \{1,\ldots,L\}$ and $q \in \{1,\ldots,Q\}$ as auxiliary variables introduced by the Lagrangian dual transform and quadratic transform, respectively. The classical FP algorithm then iteratively optimizes these variables in the following sequence
$$\underline{\boldsymbol{v}}^0\to\cdots\to\underline{\boldsymbol{v}}^{k-1}\to\underline{\boldsymbol{y}}^k\to\underline{\boldsymbol{\gamma}}^k\to\underline{\boldsymbol{v}}^k\to\cdots.$$

Specifically, the auxiliary variable $\underline{\boldsymbol{\gamma}}$ is updated by
\begin{equation}\label{update_gamma}
\gamma_{\ell q}^\star = \text{SINR}_{\ell q} = \mathbf{v}_{\ell q}^H\mathbf{H}_{\ell q,\ell}^H\mathbf{C}_{\ell q}^{-1}\mathbf{H}_{\ell q,\ell}\mathbf{v}_{\ell q}
.\end{equation}
The auxiliary variable $\underline{\boldsymbol{y}}$, which is also known as the receiver filter in WMMSE algorithm, is updated by
\begin{equation}\label{update_y}
\mathbf{y}_{\ell q}^{\star} = \left(\sigma_N^2\mathbf{I}+\sum_{i=1}^L\sum_{j=1}^Q\mathbf{H}_{\ell q,i}\mathbf{v}_{ij}\mathbf{v}_{ij}^H\mathbf{H}_{\ell q,i}^H\right)^{-1}\mathbf{H}_{\ell q,\ell}\mathbf{v}_{\ell q}.
\end{equation}
In the sequel, the beamforming vector $\underline{\boldsymbol{v}}$ is updated by
\begin{equation} \label{update_v}
\small
\begin{aligned}
\mathbf{v}^{\star}_{\ell q}(\lambda_\ell) 
&= \Bigg(\lambda_\ell\mathbf{I}+\sum_{i=1}^L\sum_{j=1}^Q\omega_{ij}(1+\gamma_{ij})\mathbf{H}_{ij,\ell}^H\mathbf{y}_{ij}\mathbf{y}_{ij}^H\mathbf{H}_{ij,\ell}\Bigg)^{-1} \\
&\quad \times \omega_{\ell q}(1+\gamma_{\ell q})\mathbf{H}_{\ell q,\ell}^{H}\mathbf{y}_{\ell q},
\end{aligned}
\end{equation}
where $\lambda_\ell$ is the Lagrange multiplier that ensures the beamforming vectors to satisfy the power constraint at BS $\ell$. The optimal $\lambda_\ell$ is determined by
\begin{equation}\label{lambda_star}
\lambda_\ell^\star = \min\left\{\lambda\geq0:\sum_{q=1}^Q\|\mathbf{v}_{\ell q}(\lambda)\|_2^2\leq P_{\ell}\right\}.
\end{equation}

In practice, $\lambda_\ell$ is typically obtained via bisection search. As indicated by \eqref{update_v} and \eqref{lambda_star}, each evaluation of $\lambda_\ell$ requires inverting an $N_{t}\times N_{t}$ matrix, which incurs substantial computational complexity and time cost. This matrix inverse operation constitutes the primary computational bottleneck of the classical FP algorithm.

\subsection{FastFP Algorithm~\cite{FastFP}}\label{AA}

To circumvent the matrix inversion required in each iteration of the classical FP algorithm, the FastFP algorithm introduces an additional nonhomogeneous transform following the Lagrangian dual transform and quadratic transform. After these three successive transforms, the objective function can be expressed as
\begin{equation}
\small
\begin{aligned}
&f_t(\underline{\boldsymbol{v}},\underline{\boldsymbol{y}},\underline{\boldsymbol{t}},\underline{\boldsymbol{\gamma}})
=\sum_{\ell=1}^{L}\sum_{q=1}^{Q}\Big[2\Re\{\omega_{\ell q}(1+\gamma_{\ell q})\mathbf{v}_{\ell q}^{H}\mathbf{H}_{\ell q,\ell}^{H}\mathbf{y}_{\ell q} \\
&\quad+\mathbf{v}_{\ell q}^{H}(\tau_{\ell}\mathbf{I}-\mathbf{D}_{\ell})\mathbf{t}_{\ell q}\}+\mathbf{t}_{\ell q}^{H}(\mathbf{D}_{\ell}-\tau_{\ell}\mathbf{I})\mathbf{t}_{\ell q}-\tau_{\ell}\mathbf{v}_{\ell q}^{H}\mathbf{v}_{\ell q} \\
&\quad- \omega_{\ell q}(1+\gamma_{\ell q})\sigma_N^{2}\mathbf{y}_{\ell q}^{H}\mathbf{y}_{\ell q}+\omega_{\ell q}\log(1+\gamma_{\ell q})-\omega_{\ell q}\gamma_{\ell q}\Big],
\end{aligned}
\end{equation}
where $\underline{\boldsymbol{t}} = \{\mathbf{t}_{\ell q}: \ell \in \{1,\ldots,L\}, q\in\{1,\ldots,Q\}\}$ is the auxiliary variable introduced by the nonhomogeneous transform. The Fast FP algorithm solves this problem through the following iterative sequence:
$$\underline{\boldsymbol{v}}^0\to\cdots\to\underline{\boldsymbol{v}}^{k-1}\to\underline{\boldsymbol{t}}^k\to\underline{\boldsymbol{y}}^k\to\underline{\boldsymbol{\gamma}}^k\to\underline{\boldsymbol{v}}^k\to\cdots.$$

The auxiliary variables $\underline{\boldsymbol{y}}$ and $\underline{\boldsymbol{\gamma}}$ are updated using \eqref{update_y} and \eqref{update_gamma}, which are identical to the classical FP algorithm. The auxiliary variable $\underline{\boldsymbol{t}}$ is updated by $\mathbf{t}^{k}_{\ell q}=\mathbf{v}^{k-1}_{\ell q}$. The beamforming vector $\underline{\boldsymbol{v}}$ is updated by
\begin{equation}
\mathbf{v}_{\ell q}^{\star}=\begin{cases}
\widetilde{\mathbf{v}}_{\ell q} & \text{if } \sum_{j=1}^Q\|\widetilde{\mathbf{v}}_{\ell j}\|_2^2\leq P_{\ell}\\
\sqrt{\frac{P_{\ell}}{\sum_{j=1}^Q\|\widetilde{\mathbf{v}}_{\ell j}\|_2^2}}\widetilde{\mathbf{v}}_{\ell q} & \text{otherwise, } 
\end{cases}
\end{equation}
where $\widetilde{\mathbf{v}}_{\ell q}=\mathbf{t}_{\ell q}+\frac{1}{\delta_{\ell}}\left(\omega_{\ell q}(1+\gamma_{\ell q})\mathbf{H}_{\ell q,\ell}^{H}\mathbf{y}_{\ell q}-\mathbf{D}_{\ell}\mathbf{t}_{\ell q}\right)$. Since the auxiliary variable $\underline{\boldsymbol{t}}$ is set to the latest $\underline{\boldsymbol{v}}$, the beamforming update essentially performs a gradient ascent step followed by a projection onto the power constraint set, with stepsize $\frac{1}{\delta_{\ell}}$. To balance computational complexity and satisfaction of the nonhomogeneous transform bound, the parameter $\delta_{\ell}$ is typically chosen as either the largest eigenvalue or the Frobenius norm of matrix $\mathbf{D}_{\ell}$.

By introducing the nonhomogeneous transform, the Fast FP algorithm eliminates the matrix inversion in the classical FP algorithm, substantially reducing computational complexity. However, due to the coarse approximation inherent in the choice of $\delta_{\ell}$, the convergence rate of the Fast FP algorithm is compromised.

\subsection{DeepFP Algorithm~\cite{DeepFP}}

To tackle the problem of choosing the parameter $\delta_{\ell}$ too conservatively in the Fast FP algorithm, the DeepFP algorithm addresses this challenge by leveraging deep unfolding to learn optimal stepsize parameters. Specifically, the DeepFP algorithm unrolls the FastFP iterations into a multi-layer feedforward deep neural network. Each layer $l$ corresponds to one iteration and maintains the same update structure as FastFP, except that the stepsize parameter $\delta_{\ell}^l$ is determined by a trainable DNN $\theta^l(\cdot)$ with the input parameters $\mathbf{t}_{\ell q}$ and $\omega_{\ell q}(1+\gamma_{\ell q}^l)\mathbf{H}_{\ell q,\ell}^{H}\mathbf{y}_{\ell q}^l - \mathbf{D}_{\ell}^l\mathbf{t}_{\ell q}$: 
\begin{equation}
\delta_{\ell}^l = \theta^l\left(\mathbf{t}_{\ell q}, \omega_{\ell q}(1+\gamma_{\ell q}^l)\mathbf{H}_{\ell q,\ell}^{H}\mathbf{y}_{\ell q}^l - \mathbf{D}_{\ell}^l\mathbf{t}_{\ell q}\right).
\end{equation}
By modeling $\delta_{\ell}$ as a function of both the current beamformer and the gradient direction, the DNN can adaptively select appropriate stepsizes across iterations.

The DeepFP model employs a hybrid training strategy. In the first stage, supervised learning minimizes the mean squared error between the model output $\underline{\boldsymbol{\widehat{v}}}$ and the label $\underline{\boldsymbol{v}}^*$ obtained from the classical FP algorithm:
\begin{equation}
\text{LOSS}_1 = \frac{1}{LQ}\sum_{\ell=1}^{L}\sum_{q=1}^{Q}\|\mathbf{\widehat{v}}_{\ell q} - \mathbf{v}_{\ell q}^*\|_2^2.
\end{equation}

In the second stage, unsupervised learning maximizes the weighted sum-rate objective directly:
\begin{equation}
\text{LOSS}_2 = -\frac{1}{LQ}\sum_{\ell=1}^{L}\sum_{q=1}^{Q}\omega_{\ell q}\mathrm{R}_{\ell q}.
\end{equation}

Compared to the Fast FP algorithm, DeepFP learns more aggressive stepsizes that accelerate convergence while maintaining stability. Nevertheless, DeepFP requires pre-defining the iteration layer number before training, and its architectural structure restricts generalization capability.

\section{Proposed GNNFP Algorithm}

\subsection{Problem Reformulation}
Unlike Fast FP and Deep FP, our proposed algorithm does not introduce the nonhomogeneous transform and the corresponding auxiliary variable $\underline{\boldsymbol{t}}$. Instead, we leverage learning-based methods to directly replace the beamforming vector update step in the classical FP algorithm.  Given fixed auxiliary variables $\underline{\boldsymbol{y}}$ and $\underline{\boldsymbol{\gamma}}$, the update of $\underline{\boldsymbol{v}}$ in the classical FP algorithm amounts to solving the following problem:
\begin{subequations}\label{problem_P_v}
\begin{align}
\min_{\underline{\boldsymbol{v}}} \quad & \sum_{\ell=1}^{L}\sum_{q=1}^{Q}\Big[\text{constant}(\mathbf{v}_{\ell q})+\mathbf{v}_{\ell q}^{\mathit{H}}\mathbf{D}_{\ell}\mathbf{v}_{\ell q} \nonumber\\
& \quad -2\Re\{\omega_{\ell q}(1+\gamma_{\ell q})\mathbf{v}_{\ell q}^{\mathit{H}}\mathbf{H}_{\ell q,\ell}^{\mathit{H}}\mathbf{y}_{\ell q}\}\Big] \\
\text{s.t.} \quad & \sum_{q=1}^Q\|\mathbf{v}_{\ell q}\|_2^2\leq \mathit{P}_\ell,\quad\ell=1,\ldots,L.
\end{align}
\end{subequations}

First, observe that the term `$\text{constant}(\mathbf{v}_{\ell q})$' is independent of $\mathbf{v}_{\ell q}$, and problem (\ref{problem_P_v}) can be decomposed independently across different cell indices $\ell$. Thus, minimizing the objective function in problem (\ref{problem_P_v}) is equivalent to solving $L$ subproblems in parallel, where the $\ell$th subproblem is formulated as:

\begin{subequations}\label{problem_P_decouple}
\small
\begin{align}
\min_{\left\{\mathbf{v}_{\ell q}\right\}_{q=1}^Q}  &\sum_{q=1}^Q\Big[\mathbf{v}_{\ell q}^\mathit{H}\mathbf{D}_\ell\mathbf{v}_{\ell q}-2\omega_{\ell q}(1+\gamma_{\ell q})\Re\{\mathbf{v}_{\ell q}^\mathit{H}\mathbf{H}_{\ell q,\ell}^\mathit{H}\mathbf{y}_{\ell q}\}\Big] \\
\text{s.t.} \quad & \sum_{q=1}^Q\|\mathbf{v}_{\ell q}\|_2^2\leq \mathit{P}_\ell.
\end{align}
\end{subequations}

Subsequently, we perform a stacking operation on the $Q$ beamforming vectors of each subproblem and construct a block-diagonal structure from the quadratic coefficient matrix $\mathbf{D}_\ell$, thereby reformulating the multi-variable optimization problem into the following single-variable formulation:  
\begin{subequations}\label{problem_single_variable}
\begin{align}
\min_{\mathbf{v}_{\ell}^{sta}}&\quad{\mathbf{v}_{\ell}^{sta}}^\mathit{H}\mathbf{D}_{\ell}^{blk}\mathbf{v}_{\ell}^{sta}-2\Re\{{\mathbf{v}_{\ell}^{sta}}^\mathit{H}\boldsymbol{b}_{\ell}\}\\
\text{s.t.}&\quad\|\mathbf{v}_{\ell}^{sta}\|_2^2\leq P_{\ell},
\end{align}
\end{subequations}
where the block diagonal matrix $\mathbf{D}^{blk}_l\triangleq\text{blkdiag}(\underbrace{\mathbf{D}_l, \mathbf{D}_l, \ldots, \mathbf{D}_l}_{Q})\in\mathbb{C}^{\mathit{N}_t\mathit{Q}\times \mathit{N}_t\mathit{Q}}$ comprises $Q$ identical matrices $\mathbf{D}_l$ arranged along the main diagonal. The stacked vector $\mathbf{v}_{\ell}^{sta}\triangleq\begin{bmatrix}\mathbf{v}_{\ell1}^\mathit{T}&\mathbf{v}_{\ell2}^\mathit{T}&\cdots&\mathbf{v}_{\ell Q}^\mathit{T}\end{bmatrix}^\mathit{T}\in\mathbb{C}^{{N}_t{Q}}$ is  obtained by vertically concatenating the $Q$ beamforming vectors. The linear coefficient vector  $\mathbf{b}_{\ell}^{sta}\in\mathbb{C}^{\mathit{N}_t\mathit{Q}}$ is constructed by stacking the weighted channel components corresponding to each user: 
$$
\mathbf{b}_{\ell}^{sta} \triangleq \begin{bmatrix}
\omega_{\ell 1}(1 + \gamma_{\ell 1})\mathbf{H}_{\ell 1,\ell}^{\mathit{H}}\mathbf{y}_{\ell 1} \\
\omega_{\ell 2}(1 + \gamma_{\ell 2})\mathbf{H}_{\ell 2,\ell}^{\mathit{H}}\mathbf{y}_{\ell 2} \\
\vdots \\
\omega_{\ell Q}(1 + \gamma_{\ell Q})\mathbf{H}_{\ell Q,\ell}^{\mathit{H}}\mathbf{y}_{\ell Q}
\end{bmatrix}.
$$

To further simplify the problem structure, we introduce a one-dimensional auxiliary constant to absorb the linear term in problem (\ref{problem_single_variable}) into the quadratic structure, yielding the following standard quadratic form:
\begin{subequations}\label{problem_final}
\begin{align}
\min_{\bar{\mathbf{v}}_{\ell}}&\quad{\bar{\mathbf{v}}_{\ell}}^\mathit{H}\mathbf{D}_{\ell}^{aug}\bar{\mathbf{v}}_{\ell}\\
\text{s.t.}&\quad\|\bar{\mathbf{v}}_{\ell_{1:\mathit{N}_t\mathit{Q}}}\|_2^2\leq P_{\ell},  \quad \bar{\mathrm{v}}_{\ell_{\mathit{N}_t\mathit{Q}+1}}=1
\end{align}
\end{subequations}
where $\bar{\mathbf{v}}_{\ell_{1:\mathit{N}_t\mathit{Q}}}$ comprises the first $\mathit{N}_t\mathit{Q}$ elements of $\bar{\mathbf{v}}_{\ell}$. The augmented matrix $\mathbf{D}_{\ell}^{aug}\in\mathbb{C}^{({\mathit{N}_t\mathit{Q}}+1)\times ({\mathit{N}_t\mathit{Q}}+1)}$ is defined as  
$$\mathbf{D}_{\ell}^{aug}\triangleq\begin{bmatrix}\mathbf{D}^{blk}_\ell&-\mathbf{b}_{\ell}^{sta}\\-{\mathbf{b}_{\ell}^{sta}}^\mathit{H}&0\end{bmatrix},$$ and the augmented vector $\bar{\mathbf{v}}_{\ell}\in\mathbb{C}^{({\mathit{N}_t\mathit{Q}}+1)\times 1}$ is constructed as $\bar{\mathbf{v}}_{\ell}=\begin{bmatrix}{\mathbf{v}_{\ell}^{sta}}^\mathit{T}&1\end{bmatrix}^\mathit{T}$.

Through the above equivalent transformations, the original problem (\ref{problem_P_v}) has been reformulated into a concise standard quadratic form (\ref{problem_final}) that is particularly amenable to the training framework of GNNs. 

\subsection{GNN Structure Design}
In this section, we present the detailed design of our GNN architecture tailored for solving the single-variable standard quadratic problem (\ref{problem_final}). Our design follows a problem-aware paradigm. At a high level, the graph topology is constructed to directly reflect the mathematical structure of the quadratic form ${\bar{\mathbf{v}}_{\ell}}^\mathit{H}\mathbf{D}_{\ell}^{aug}{\bar{\mathbf{v}}_{\ell}}$. Built upon this structure, the GNN model learns variable interactions through a multi-layer message passing mechanism and enforces the power constraint (\ref{problem_final}b) via a differentiable projection layer. After forward propagation, the graph nodes output the predicted optimization variable ${\bar{\mathbf{v}}_{\ell}}^{\star}$. The model is trained in an unsupervised manner by directly minimizing the objective function (\ref{problem_final}a). The key components of the architecture are detailed below:

$\bullet$ \textbf{Graph Construction Strategy.} We construct a fully connected graph  $\mathcal{G}=(\mathcal{V},\mathcal{E})$ for each decoupled subproblem, where the mathematical structure of problem (\ref{problem_final}) is directly embedded into the graph topology and features.  The node set $\mathcal{V}$ consists of $\mathit{N}_{var}=\mathit{N}_t\times\mathit{Q}$ variable nodes corresponding to the entries of  $\bar{\mathbf{v}}_{\ell_{1:\mathit{N}_t\mathit{Q}}}$, plus one additional constant node representing the augmented dimension with a fixed value $\bar{\mathrm{v}}_{\ell_{\mathit{N}_t\mathit{Q}+1}}=1$.  

This fully connected topology is motivated by the fact that the quadratic form ${\bar{\mathbf{v}}_{\ell}}^\mathit{H}\mathbf{D}_{\ell}^{aug}{\bar{\mathbf{v}}_{\ell}}$ couples all variable pairs through the coefficient matrix $\mathbf{D}_{\ell}^{aug}$. Thus, the node features are extracted from the diagonal elements 
$$\mathbf{x}_i=\Big[\Re\{\mathbf{D}_{\ell_{ii}}^{aug}\},\Im\{\mathbf{D}_{\ell_{ii}}^{aug}\}\Big]\in\mathbb{R}^2,$$
which encode the self-energy terms, whilst the edge features capture the pairwise coupling strength
$$\mathbf{e}_{ij}=\Big[\Re\{\mathbf{D}_{\ell_{ij}}^{aug}\},\Im\{\mathbf{D}_{\ell_{ij}}^{aug}\}\Big]\in\mathbb{R}^2.$$ 
$\bullet$ \textbf{Feature Embedding Layer.}  The raw 2-dimensional features are embedded into a higher-dimensional latent space to enhance representational capacity. Specifically, we employ two separate Multi-Layer Perceptrons (MLPs) to independently encode node and edge features: 
\begin{equation}\mathbf{h}_i^{(0)}=\mathrm{MLP}_{\mathcal{V}}(\mathbf{x}_i)\in\mathbb{R}^{m},\quad\mathbf{e}_{ij}^{(0)}=\mathrm{MLP}_{\mathcal{E}}(\mathbf{e}_{ij})\in\mathbb{R}^{n},\end{equation}
incorporating ReLU activations for non-linearity, batch normalization for training stability, and dropout for regularization.

$\bullet$ \textbf{EdgeConv Message Passing Layers.}  The core of our architecture consists of $K$ stacked EdgeConv layers that iteratively refine node representations via message passing. At layer $k$, the message from node $j$ to node $i$ is computed as:  
\begin{equation}\mathbf{m}_{ij}^{(k)}=\mathrm{MLP}^{(k)}\bigg(\Big[\mathbf{h}_i^{(k-1)}\|\mathbf{h}_j^{(k-1)}\|\mathbf{e}_{ij}^{(0)}\Big]\bigg),\end{equation}
where $||$ denotes concatenation. By incorporating $\mathbf{D}_{\ell}^{aug}$ as edge features, this design directly injects the structure of the problem (\ref{problem_final}) into message computation. 
Then, messages are aggregated using a dual max-mean pooling strategy that combines two complementary operations:
\begin{equation}\mathbf{h}_i^{(k)}=\left[\max_{j\in\mathcal{N}(i)}\mathbf{m}_{ij}^{(k)}\parallel\frac{1}{|\mathcal{N}(i)|}\sum_{j\in\mathcal{N}(i)}\mathbf{m}_{ij}^{(k)}\right],\end{equation}
where ${\mathcal{N}(i)}$ denotes the neighborhood of node $i$. Notably, max-pooling identifies dominant coupling patterns, while mean-pooling provides robust global statistics. The concatenation of both pooling results yields more expressive node representations.  

$\bullet$ \textbf{Output Decoder and Constraint Projection.}  After $K$ EdgeConv layers, the node representations are concatenated with the initial embeddings to form: 
\begin{equation}\mathbf{h}_i^\mathrm{final}=[\mathbf{h}_i^{(K)}\|\mathbf{h}_i^{(0)}].\end{equation}
This skip connection preserves low-level geometric features from the original problem structure and mitigates gradient vanishing during training.  For the $\mathit{N}_{var}$ variable nodes, a linear decoder predicts the real and imaginary components as
\begin{equation}\Big[\Re\{\hat{\mathrm{v}}_{\ell_{i}}^{sta}\},\Im\{\hat{\mathrm{v}}_{\ell_{i}}^{sta}\}\Big]=\mathbf{W}_{\mathrm{dec}}\mathbf{h}_i^{\mathrm{final}}+\mathbf{b}_{\mathrm{dec}}\in\mathbb{R}^2.\end{equation}
which are combined to form the complex vector $\hat{\mathbf{v}}_{\ell}^{sta}\in\mathbb{C}^{N_tQ}$. To ensure strict satisfaction of the power constraint in (\ref{problem_final}b), we apply a differentiable normalization layer: 
\begin{equation}
{\mathbf{v}_\ell^{sta}}^{\star}=\begin{cases}\quad
\widehat{\mathbf{v}}_{\ell}^{sta} & \text{if } \quad\|\widehat{\mathbf{v}}_{\ell}^{sta}\|_2^2\leq P_{\ell}\\
\sqrt{\frac{P_{\ell}}{\|\widehat{\mathbf{v}}_{\ell}^{sta}\|_2^2+\epsilon}}\cdot\widehat{\mathbf{v}}_{\ell}^{sta} & \text{if } \quad\|\widehat{\mathbf{v}}_{\ell}^{sta}\|_2^2> P_{\ell},
\end{cases}
\end{equation}
where $\epsilon$ is a tiny bias to prevent numerical instability. This projection maintains end-to-end differentiability with respect to model parameters.  Final forward propagation output ${\bar{\mathbf{v}}_{\ell}}^{\star}=\big[{{\mathbf{v}_\ell^{sta}}^{\star T}} \; 1\big]$ is the concatenation of the differentiable normalization layer output and the constant node output. The complete process enables gradient-based unsupervised training that directly minimizes the objective function without requiring labeled optimal solutions. 

From an intelligent algorithmic perspective, the proposed architecture integrates three synergistic design principles. First, structure-aware graph construction embeds the quadratic form (\ref{problem_final}a) into graph topology, wherein nodes represent variables and edges encode coupling coefficients from the augment matrix $\mathbf{D}_{\ell}^{aug}$. Second, edge-aware message passing exploits these coupling coefficients to guide information aggregation, enabling the model to learn interaction patterns critical for optimization. Third, differentiable projection enforces the power constraint (\ref{problem_final}b) while preserving gradient flow, thereby eliminating post-processing requirements. 

\section{Experimental Results}
\subsection{Experimental Setup}
We generate channel data for the multi-cell network under the same setting as in \cite{FPPart1,FastFP,DeepFP}. Specifically, we consider a 7-hexagonal-cell network where each BS is located at the cell center and serves $Q=6$ randomly distributed downlink users. The BSs and users are equipped with $N_t=8$ and $N_r=2$ antennas, respectively. For simplicity, all users are assigned equal weights. The inter-BS distance is $D = 2R = 0.8$ km, and the cell radius is $R/\sqrt3$. The maximum transmit power and background noise power are set to 20 dBm and -90 dBm, respectively. The downlink path loss is modeled as $128.1+37.6\log_{10}r+\tau$ (in dB), where $r$ is the BS-to-user distance in kilometers, and $\tau$ is a zero-mean Gaussian random variable with 8 dB standard deviation to capture shadowing effects.

The proposed GNN model is trained on a dataset comprising 3,000 network samples. The dataset is randomly partitioned into training, validation, and test sets (70:15:15). Node and edge feature encoders use MLPs with dimensions 2$\rightarrow$16$\rightarrow$ 8. There are $K=3$ stacked EdgeConv layers (MLP: 24$\rightarrow$32$\rightarrow$16, 72$\rightarrow$32$\rightarrow$16, and 72$\rightarrow$32$\rightarrow$16) employed, with aggregated outputs of 32 dimensions each. Dropout rates of 0.1 (MLP layers) and 0.2 (before decoder) are applied, with batch normalization used in all MLP layers.  

\subsection{WSR Comparison with Other FP-based Algorithms}
\begin{figure}[htbp] 
  \centering 
  \includegraphics[scale=0.4]{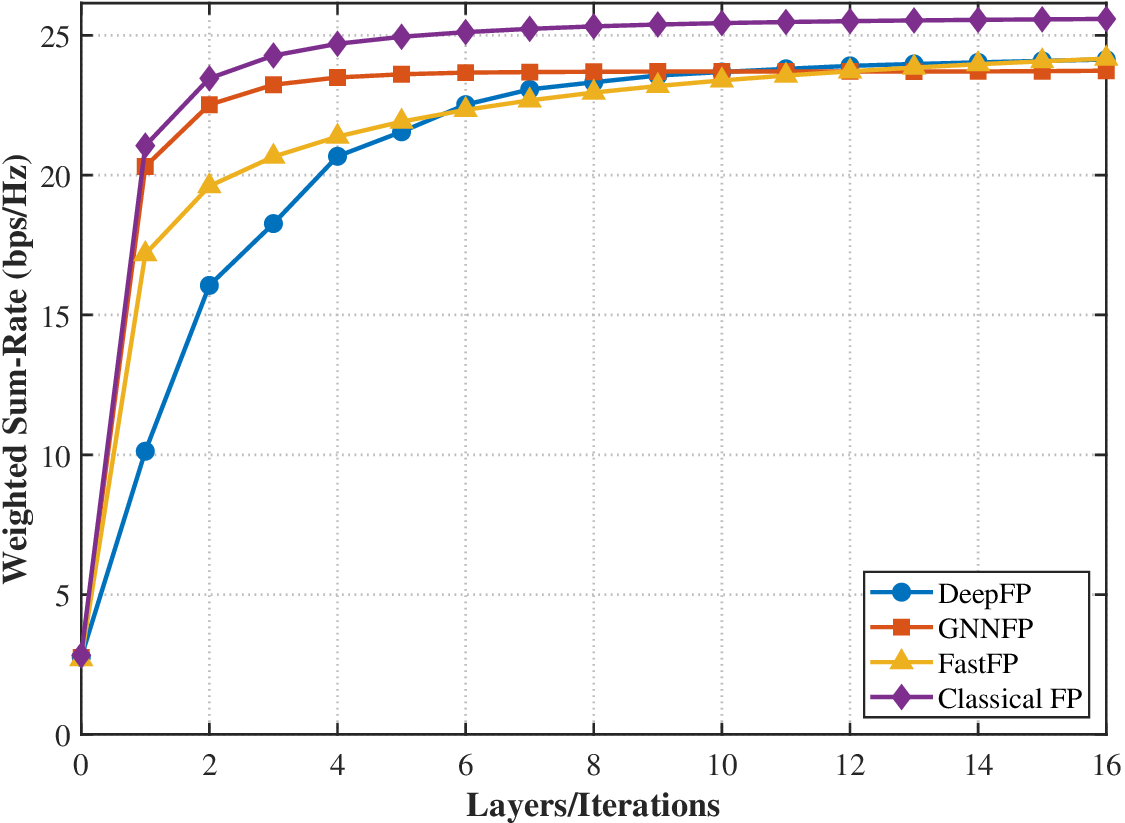} 
  \caption{The weighted Sum-Rate performance of 4 FP-Based algorithms} 
  \label{fig:sum_rate} 
\end{figure}

Figure \ref{fig:sum_rate} illustrates the weighted sum-rate performance comparison of four FP-based algorithms as a function of the number of layers/iterations. 
We train and evaluate the DeepFP algorithm as described in \cite{DeepFP} using its publicly accessible source code. We employ 5,000 network samples for DeepFP training, which are more than those used for GNNFP. 
It is worth noting that DeepFP requires pre-defining the number of iteration layers during training, which is set to 16 in this sub-experiment.
Since the beamforming vectors $\underline{\boldsymbol{v}}$ are updated via closed-form solutions in each iteration, which is approximately optimal, the classical FP algorithm demonstrates superior asymptotic performance, thereby establishing the performance upper bound.
Our proposed GNNFP algorithm exhibits remarkably rapid convergence behavior, achieving approximately 99\% of its converged value (obtained at 16 iterations) within merely 5 iterations. 
Furthermore, our proposed algorithm significantly outperforms both FastFP and DeepFP for iteration numbers up to 6, with only a marginal performance gap compared to the classical FP algorithm. When the number of iterations exceeds 10, its performance becomes comparable to that of FastFP and DeepFP.

\subsection{Computational and Model Size Evaluation}
As mentioned in subsection V.B, DeepFP employs distinct model parameters for computing the stepsize $\delta$ at each layer, which necessitates pre-defining the number of iteration layers before training. That is, each layer of DeepFP requires distinct parameters across layers. In contrast, GNNFP trains a unique model for the quadratic programming in (\ref{problem_P_v}), which is universally applicable to the problem of this type.
As such, model parameters in GNNFP are shared across layers, i.e., all unfolded layers reuse the same set of parameters, thereby facilitating learning to unfold an arbitrary number of iterations.
\begin{table}[htbp]
\centering
\caption{Computational Performance and Model Size comparison between DeepFP and GNNFP}
\label{modelsize}
\footnotesize
\setlength{\tabcolsep}{3pt}
\renewcommand{\arraystretch}{1.1}
\begin{tabular}{|l|@{\hspace{3.1pt}}c@{\hspace{3.1pt}}|@{\hspace{3.1pt}}c@{\hspace{3.1pt}}|@{\hspace{3.1pt}}c@{\hspace{3.1pt}}|}
\hline
& \textbf{Time/layer} & \textbf{Model size} & \textbf{Parameters} \\
\hline
DeepFP (5 layers) & 0.3571 sec & 0.74 MB & 47365 = 9473$\times$5 \\
\hline
GNNFP & 0.2577 sec & 0.05 MB & 7890 \\
\hline
\end{tabular}
\end{table}

Table \ref{modelsize} presents a comparison of computational time cost and model size between the GNNFP and DeepFP algorithms. 
 We take a 5-layer DeepFP model for comparison with our proposed GNNFP algorithm, as their sum-rate performance is comparable (as demonstrated in Table \ref{generalize performance}). Remarkably, GNNFP contains significantly fewer parameters than the 5-layer DeepFP. Even for a single unfolded layer, the former has fewer parameters and takes less time than the latter. Since fewer model parameters usually require fewer training samples, this explains why fewer samples were needed to train GNNFP compared to DeepFP.
The third column in the table indicates the substantially smaller storage requirement of GNNFP compared with DeepFP. 
To summarize, the proposed GNNFP algorithm outperforms DeepFP in terms of inference time, model size, and the number of parameters, serving as a more favorable candidate in practical systems.
\subsection{Generalization Performance}

\begin{table}[htbp]
\centering
\caption{Generalization performance for different numbers of users per cell ($Q=3,\dots,8$)}
\label{generalize performance}
\footnotesize
\setlength{\tabcolsep}{3.15pt}
\renewcommand{\arraystretch}{1.1}
\begin{tabular}{|l|c|c|c|c|c|c|}
\hline
& \textbf{$Q=3$} & \textbf{$Q=4$} & \textbf{$Q=5$} & \textbf{$Q=6$} & \textbf{$Q=7$} & \textbf{$Q=8$} \\
\hline
FP (iters = 100)& 100 & 100 & 100 & 100 & 100 & 100 \\
\hline
FP (iters = 5) & 96.52 & 96.62 & 96.88 & 96.88 & 96.89 & 96.80 \\
\hline
FastFP (iters = 5) & 85.10 & 84.96 & 84.96 & 84.42 & 83.84 & 84.76 \\
\hline
DeepFP (layers = 5)& 91.01 & 88.54 & 88.20 & 91.12 & \textbf{N/A} & \textbf{N/A} \\
\hline
GNNFP (layers = 5)& 93.92 & 93.22 & 93.05 & 91.61 & 89.68 & 86.85 \\
\hline
\end{tabular}
\end{table}

We further evaluate the generalization capability of the proposed algorithm across different numbers of users per cell $Q$. Table~\ref{generalize performance} presents the weighted sum rate percentages for four FP-based algorithms after 5 iterations (layers) normalized against the classical FP with 100 iterations as the baseline. Both GNNFP and DeepFP algorithms were trained with $Q=6$. The number of unfolding layers in DeepFP was pre-defined as 5 before training. Owing to permutation equivariance of GNN architectures, our proposed GNNFP algorithm learns the interaction patterns between information entities, thus enabling upward generalization in $Q$, which is not applicable in DeepFP. Moreover, GNNFP achieves superior performance to FastFP as the number of iterations/layers increases. Notably, our proposed algorithm exhibits remarkable performance improvement during downward generalization in $Q$, which is likely attributed to the reduced problem dimensionality.

\section{Conclusion}
This paper proposed a lightweight and generalizable FP-based algorithm, named GNNFP, for the beamforming optimization in multi-cell MU-MIMO systems targeting weighted sum-rate maximization.
Built upon the classical FP framework, our approach innovatively transforms the beamforming update subproblem into a standard quadratic form and employs a structure-aware GNN to directly solve the reformulated problem.
Experimental results validate the superiority of GNNFP in terms of convergence speed, computational efficiency, model compactness, and generalization capability.

\end{document}